\documentclass[reprint,showpacs,amsmath,amssymb,aps,prl]{revtex4-1}
\usepackage{graphicx}% Include figure files
\usepackage{dcolumn}% Align table columns on decimal point
\usepackage{bm}
\usepackage[utf8]{inputenc}

\begin{document}

\preprint{APS/123-QED}
\title{Magnetic topology and current channels in plasmas with \\toroidal current density inversions}
\author{D. Ciro}
 \email{davidcirotaborda@gmail.com}
\author{I. L. Caldas}
 \email{ibere@if.usp.br}
\affiliation{Departamento de Física Aplicada, Universidade de São Paulo, 05508-090, São Paulo, Brazil.}
%\date{\today}

\begin{abstract}
The equilibrium magnetic field inside axisymmetric plasmas with inversions on the toroidal current density is studied. Structurally stable non-nested magnetic surfaces are considered. For any inversion in the internal current density the magnetic families define several positive current channels about a central negative one. A general expression relating the positive and negative currents is derived in terms of a topological anisotropy parameter. Next, an analytical local solution for the poloidal magnetic flux is derived and shown compatible with current hollow magnetic pitch measurements shown in the literature. Finally, the analytical solution exhibits non-nested magnetic families with positive anisotropy, indicating that the current inside the positive channels have at least twice the magnitude of the central one.

\end{abstract}

\pacs{52.55.-s, 52.30.Cv, 41.20.Gz}
\maketitle

In tokamak devices a toroidal magnetic field confine the orbits of the charged particles inside a chamber. A toroidal current flowing within the plasma produces a complementary magnetic field preventing particle drift looses. Non-inductive drive mechanisms help to sustain this current for long pulses with small or negative inductive drive and during a slow transition from positive to negative toroidal current~\cite{jet92}. A relevant question arising from this situations is that of the structure of the magnetic field if the toroidal current density becomes negative in some region of the plasma. In the last decade, the achievement of quasi-steady-state alternating current scenarios~\cite{HT706} and the observation of stiff structures with nearly zero magnetic pitch angles in a finite region about the plasma center~\cite{hawkes01,fujita05} has attracted attention to the problem of current reversal equilibrium configurations (CRECs).

In this work we study the equilibrium topology of the magnetic field subjected to azimuthal (toroidal) current density inversions in an axisymmetric plasma. In the following pages it is shown that the equilibrium topology is composed of non-nested families of nested magnetic surfaces, where each magnetic family defines a current channel inside the plasma. The relation between the currents in the channels is studied in terms of topological quantities. In a more quantitative approach, an analytical solution of the equilibrium problem about a region of interest provides the topology of the magnetic surfaces and the relevant control parameters as well as their bifurcation values controlling the transition between different equilibrium configurations. This is done without specifying further plasma profiles or arbitrary functions. The obtained solution agrees with several published equilibria~\cite{martynov03,wang04,bizarro05,caroline11} while providing a simple understanding of the control parameters and their 
relation with the current in the different channels.

For axisymmetric systems the equilibrium magnetic field may be written
\begin{equation}\label{01}
\bm B=\nabla\psi\times\nabla\phi+F\nabla\phi,
\end{equation}
where $\psi(R,z)=RA_\phi(R,z)$ and $F(R,z)=RB_\phi(R,z)$ are proportional to the poloidal magnetic flux and current respectively. $B_\phi(R,z)$ and $A_\phi(R,z)$ are the azimuthal components of the magnetic field and vector potential and $(R,\phi,z)$ the usual cylindrical coordinates. The term $\nabla\psi\times\nabla\phi$ corresponds to the poloidal magnetic field $\bm B_p$, i.e. the projection of the magnetic field over a plane $\phi=const.$ It is clear that $\bm B\cdot\nabla\psi=0$, so that the magnetic field lines remain attached to the level sets of $\psi(R,z)$ called {\it magnetic surfaces}.

For axisymmetric magnetic confinement devices the main source of poloidal field is the plasma current, consequently, the level sets of $\psi(R,z)$ form families of nested tori inside the plasma. In equilibrium, the toroidal current flowing inside a magnetic surface labeled by $\psi$ is given by the Ampère's law as
\begin{equation}\label{03}
 \mu_0I_t(\psi)=\oint_{\Gamma_\psi} \bm B\cdot d\bm l=\bm\phi\cdot\oint_{\Gamma_\psi}\nabla\psi\times d\bm l=\pm\oint_{\Gamma_\psi} |\nabla\psi| dl,
\end{equation}
where $\Gamma_\psi$ is a {\it magnetic circuit} resulting from the intersection of the magnetic surface $\psi$ with an arbitrary plane $\phi=const.$ For a simple magnetic circuit the product $\nabla\psi\times d\bm l$ always points in the same direction, the current is negative when $\nabla\psi$ is inwards and positive otherwise. The existence of current density reversals suggest that some magnetic surface $\psi_0$ contains a vanishing current; from~(\ref{03}) this requires $|\nabla\psi|=0$ along the continuous magnetic circuit $\Gamma_0=\Gamma_{\psi_0}$. This requires that for {\it any} coordinate system $\{u,v\}$, the equation $\partial_u\psi(u,v)=0$ leads to the same relation between $u$ and $v$ that the equation $\partial_v\psi(u,v)=0$. In addition, such relation must satisfy $\psi(u,v)=\psi_0$. This kind of degeneracy is possible for one-dimensional problems, but leads to structural instability in general two-dimensional equilibria~\cite{bizarro05}, i.e. any variation in the poloidal field, will destroy the 
topology of the surface $\psi=\psi_0$.

A more feasible situation, i.e. structurally stable, requires that $\nabla\psi$ vanishes at isolated points inside the plasma. Some of these points must be saddle, each one introducing four branches where the condition $\psi(R,z)=\psi_c$ is satisfied, with $\psi_c$ the value of the poloidal flux on the critical point. This guarantees that the poloidal field reverses over the surface $\psi(R,z)=\psi_c$ after vanishing at each saddle.

\begin{figure}[h]
 \centering
 \includegraphics[width=0.45\textwidth]{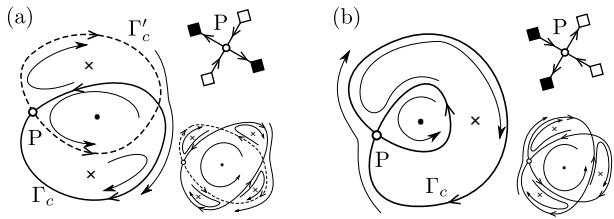}
 \caption{\label{f1}Odd and even systems of axisymmetric islands. Corresponding diamonds are connected by the separatrix $\Gamma_c$. The arrows indicate the direction of $\bm B_p$ and "$\cdot$" or "$\times$" in the magnetic axes indicate the current direction for each channel.}
\end{figure}

The CRECs topologies may be classified depending on how the separatrix $\Gamma_c=\Gamma_{\psi_c},$ connects two branches of the same hyperbolic point~\footnote{The separatrix must be followed so that the trajectory is differentiable}. If $\Gamma_c$ connects two opposite branches of a saddle the resulting circuit is {\it simple} but {\it non-unique}, i.e. there is a second simple circuit $\Gamma_c'$ connecting the remaining two branches; this leads to an even number of axisymmetric islands (Fig.~\ref{f1}a). On the other hand, when $\Gamma_c$ connects two non-opposite branches the circuit is {\it non-simple} but {\it unique} and leads to an odd number of islands (Fig.~\ref{f1}b). Separatrixes delimit several families of nested magnetic surfaces acting as {\it current channels} inside the plasma. These axisymmetric channels are also called {\it non-nested surfaces} in the literature. In Fig.~\ref{f1} the poloidal field direction reveals the sign of the toroidal current inside each channel. In summary, any internal toroidal current density reversal in a two-dimensional magnetic equilibrium leads to a chain of axisymmetric islands with positive current about a central structure with negative current.

\begin{figure}[h]
 \centering
 \includegraphics[width=0.45\textwidth]{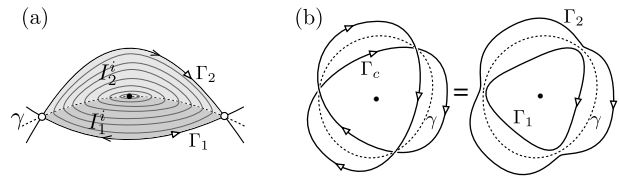}
 \caption{\label{f2}(a)~Regions for the {\it half-currents} between $\gamma$ and the separatrix. (b)~Decomposition of a self-intersecting circuit into the simple circuits $\Gamma_1$ and $\Gamma_2$. Open arrows show the circuits orientations and simple ones the poloidal field direction.}
\end{figure}

As the poloidal field only vanishes at isolated points and never reverses in a regular magnetic surface, its line integral on a magnetic circuit can never be zero. From this, the current enclosed by a magnetic circuit just outside the positive channels is positive, in consequence, the total current in the positive channels must exceed the current of the negative one, not only equate its magnitude. Each positive channel contains an elliptic point and is bounded by a separatrix and two hyperbolic points~\footnote{This may be generalized to include several elliptic points and some internal separatrix}. We can always build a curve $\gamma$ parallel to $\nabla\psi$ that passes through all the hyperbolic and elliptic points defining the positive current channels (Fig.~\ref{f2}a). The circuit $\gamma$ encloses a vanishing current, given that $\bm B_p\cdot d\bm l=0$ in all its points. By decomposing the critic magnetic circuit $\Gamma_c$ into two simple circuits $\{\Gamma_1,\Gamma_2\}$ (Fig.~\ref{f2}b) it can be shown that
\begin{equation}
 -\oint_{\Gamma_1}B_p dl=\sum_i I_1^i=I_1\mbox{ , }\oint_{\Gamma_2}B_p dl=\sum_i I_2^i=I_2,
\end{equation}
where $I_{1,2}^i$ is the current flowing through the region between $\gamma$ and $\Gamma_{1,2}$ on the $i'th$ positive current channel. The relative difference between the {\it half currents} on the $i$'th channel is $\eta_i=(I_2^i-I_1^i)/I^i$, with $I^i=I_1^i+I_2^i$ the total current inside that channel. The current flowing inside all the positive channels is $I_+=I_1+I_2$ and the one flowing in the central channel is $I_-=-I_1$. Using the previous relations we have
\begin{equation}\label{04}
 I_+=\frac{2}{1-\eta}|I_-|,
\end{equation}
with $\eta=\sum_i\eta_iI_i/\sum_i I_i$. Equation~(\ref{04}) relates the total current in the positive channels with the negative current in the central one through the parameter $\eta$, that measures the anisotropy of the positive current channels. For small anisotropies we have $I_+\sim 2I_-$ and in simple cases where the islands are created in regions with monotonic variation of the current density we expect $\eta>0$, leading to $I_+>2|I_-|$.

To study the properties of the equilibrium solutions, like the size or number of islands, we put $\eta$ in terms of geometric parameters. For this, we develop a {\it local} successive approximations method that account for the non-nested topologies while keeping a simplified physical picture of the equilibrium.

From (\ref{01}) and $\nabla\times\bm B=\mu_0 \bm j$ it can be verified that
\begin{equation}\label{07}
 R\nabla\cdot(R^{-2}\nabla\psi)=-\mu_0 j_{\phi}.
\end{equation}
Also, from the single-fluid MHD equilibrium~\cite{freidberg82} the force balance condition $\nabla p=\bm j\times\bm B$, leads to the relation
\begin{equation}\label{08}
 \mu_0 j_\phi=\mu_0 Rp'(\psi)+R^{-1}F(\psi)F'(\psi),
\end{equation}
where the prime denotes $d/d\psi$, $p(\psi)$ is the kinetic pressure and $F(\psi)$ is defined after~(\ref{01}). Equating (\ref{07}) and (\ref{08}) gives the well known Grad-Shafranov (G-S) equation~\cite{shafranov60,grad58}, a nonlinear elliptic partial differential equation. In (\ref{08}), self-consistent {\it surface functions} $\{p(\psi),F(\psi)\}$ and boundary conditions characterize the magnetohydrodynamic equilibrium reached by the system~\cite{freidberg82}. In regular situations the equilibrium is described by a single family of magnetic surfaces, requiring a single choice of $\{p(\psi),F(\psi)\}$ and a single boundary (the plasma edge). This may not be appropriate for CRECs, but in a first approach a single choice of the sources $p(\psi)$ and $F(\psi)$ can present current density inversions and multiple magnetic families~\cite{martynov03,wang04}. Other works uses successive approximations to the solution with prescribed zero-order current density models and boundary conditions~\cite{bizarro05,caroline11}.

Formally, a broad range of choices for the arbitrary functions $p(\psi)$ and $F(\psi)$ or zero-order profiles may lead to current reversals and non-nested configurations, but the underlying description of the equilibrium topology is not restricted to the particularly chosen model. In the following we assume directly the existence of a small negative minimum of the current density in some small region of the plasma. With this approach we cover a wide range of arbitrary choices without specifying the global configuration of the plasma.

\begin{figure}[h]
 \centering
 \includegraphics[width=0.45\textwidth]{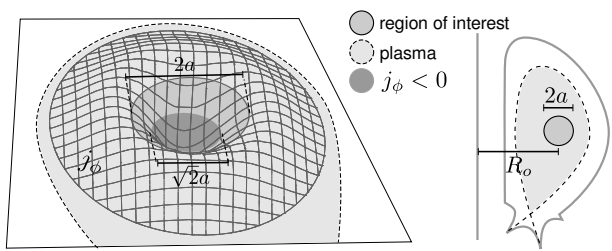}
 \caption{\label{f3} Representation of the {\it region of interest}.}
\end{figure}

Near a current density minimum at $\bm r_0$, a usual Taylor expansion gives $j_\phi(\bm r)\approx j_\phi(\bm r_0)+1/2(\delta \bm r\cdot\nabla)^2_{\bm r_0}j_\phi(\bm r)+O(\delta\bm r^3)$. For an up-down symmetric equilibrium this can be casted like $j_\phi=j_0+\iota (r^2-\kappa r^2\cos^2\theta)$, where $(r,\theta)$ are local polar coordinates centered at $\bm r_0$. Here, $j_0=j(\bm r_0)<0$, $r=|\bm r-\bm r_0|$ and $\theta$ is measured counterclockwise between $\bm e_R$ and $\bm r-\bm r_0$. The parameters $\iota$ and $\kappa$ are related to the curvature and ellipticity of $j(\bm r)$ about $\bm r_0$. For $\kappa=0$, a circle with radius $a=\sqrt{-2j_0/\iota}$ contains a zero current and for $r<a/\sqrt{2}$ we have $j_\phi<0$. Since we are interested in the topology of the magnetic surfaces in a region containing an almost vanishing toroidal current we define a {\it region of interest} with length $2a$, centered at $\bm r_0$ (Fig.\ref{f3}).

Provided that the region of interest is small we can solve (\ref{07}) by a {\it local} scheme analogous to the successive approximations method~\cite{freidberg82}. In our case the inverse aspect ratio, $\epsilon=a/R_0$, is a reliably small parameter defined from the region of interest instead of the plasma radius, in consequence, a fast convergence of the approximations is expected. Since the local current description is not intended to provide the vanishing of the current density at the plasma edge, the following solution is valid for $r\lesssim a$.

Defining $R=R_0+ax$ and $z=ay$ we can write~(\ref{07}) about the current minimum like
\begin{equation}\label{09}
 \left(\partial^2_x+\partial^2_y-\frac{\epsilon}{1+\epsilon x}\partial_x\right)\psi=(1+\epsilon x)(1-2r^2+2\kappa x^2),
\end{equation}
with $r$ measured in units of $a$ and $\psi$ in units of $\mu_0 |j_0|a^2R_0$. The parameter $\kappa$ corresponds to the squared eccentricity of the current density level sets. The solutions of (\ref{09}) will depend on two parameters only, allowing a detailed study of the bifurcations that change the topology of the equilibrium in the region of interest.

\begin{figure}[h]
 \centering
 \includegraphics[width=0.45\textwidth]{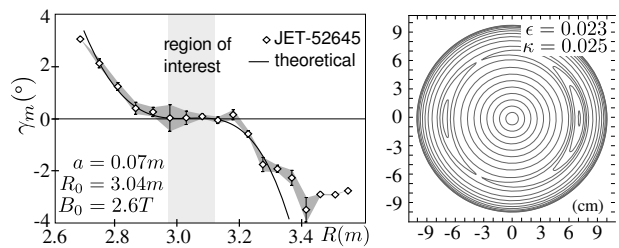}
 \caption{\label{f4}Reproduction of experimental magnetic pitch angles in a current hole and reconstruction of the magnetic surfaces. Experimental data is from N.C. Hawkees {\it et al.}, Phys. Rev. Lett. 87, 115001 (2001).}
\end{figure}

Assuming a large aspect ratio $\epsilon<<1$ and small ellipticity $|\kappa|<<1$, we write the nondimensional flux as $\psi(r,\theta)=\psi_0(r)+\epsilon\psi_1(r,x)+O(\epsilon^2)$. Excluding the ellipticity in the zero-order calculations we have $\psi_0(r)=(1-r^2/2)r^2/4$, consequently, the first order problem becomes
\begin{equation}\label{10}
 \Delta\psi_1=\left(3-5r^2\right)\frac{x}{2}+\frac{2\kappa}{\epsilon}x^2,
\end{equation}
with $\Delta$ the Laplace operator for a plane. Defining $\psi_1(r,x)=\bar{\psi_1}(r,x)+\kappa x^4/6\epsilon$ we have $\Delta\bar{\psi_1}=\left(3-5r^2\right)x/2$. Then, we reduce the problem to a simple ODE by introducing the ansatz $\bar{\psi_1}=f(r)x$. After straightforward integrations and substitutions we obtain the poloidal flux function to a first order in $\epsilon$.
\begin{equation}\label{11}
 \psi(r,x)=\left(1-\frac{1}{2}r^2\right)\frac{r^2}{4}+\epsilon x\left(1-\frac{5}{9}r^2\right)\frac{3r^2}{16}+\frac{\kappa}{6}x^4.
\end{equation}
With this local solution we can study the magnetic field in the region of interest $r\lesssim O(1)$. For instance, if we write the toroidal field as $B_t=B_0/(1+\epsilon x)$, the magnetic pitch angle, defined by $\gamma_m=\arctan(B_p/B_t)$, becomes $\gamma_m=\arctan(\tau|\nabla\psi|)$, with $\tau=\mu_0a|j_0|/B_0$. By a suitable choice of the region of interest and adjust of $\kappa$ we can fit experimental measurements of $\gamma_m$ about a current minimum and model the corresponding magnetic surfaces (Fig.~\ref{f4}).
This local solution agrees well with the reported Motional Stark Effect measurements in a current hole in~\cite{hawkes01} and predicts $j_0\sim-42.6$~kA/m$^2$, a value within the error reported by Hawkees {\it et al.}

\begin{figure}[h]
 \centering
 \includegraphics[width=0.45\textwidth]{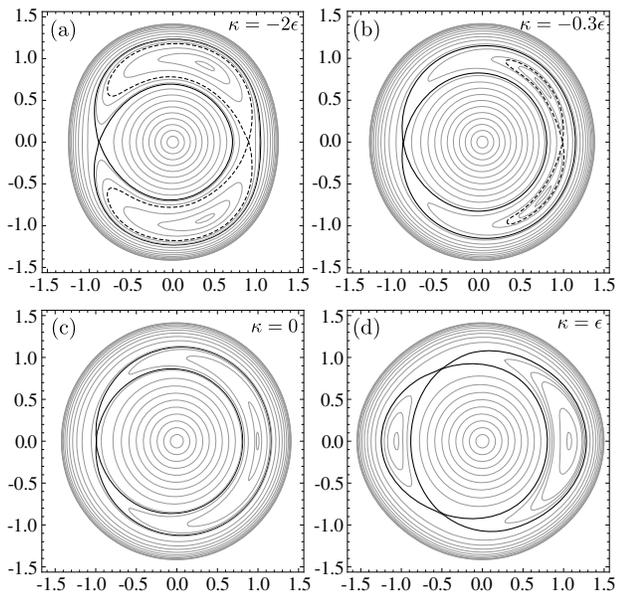}
 \caption{\label{f5} Levels sets of $\psi$ from (\ref{09}) with $\epsilon=0.1$. The continuous separatrix defines the current channels and the dashed one the island internal structure.}
\end{figure}

The structure of the magnetic surfaces depend on the relative positions of the critical points ($\nabla\psi=0$), their type (elliptic or hyperbolic) and the value of $\psi$ on them. Since $r$ and $x$ are independent coordinates for $\theta\neq \{0,\pi\}$, the off-axis critical points satisfy $\partial_r\psi=\partial_x\psi=0$.
The origin $r=0$, is always an elliptic point between two critical points in the $x$-axis with $x_c\sim\pm1$. The type of the these points depend on the ellipticity $\kappa$. To understand this, we use the equation resulting of $\partial_r\psi=0$ and the condition for off-axis critical points $|x_c|<r_c$. Taking into account the domain of the first term in~(\ref{11}) we obtain $|\kappa|\gtrsim\epsilon/8$. The increase of the ellipticity leads to bifurcations about $|\kappa|\sim\epsilon/8$, where off-axis critical points are created from those in the $x$-axis that change of type after the bifurcation occurs. A negative (positive) ellipticity represents a vertically (horizontally) elongated current density.  For $\kappa<\epsilon/8$, there is a single positive current channel about the negative one (Fig.~\ref{f5}a-c). For $\kappa\lesssim-\epsilon/8$ the positive channel contains an internal separatrix (Fig.~\ref{f5}a,b) that for large anisotropies (or small $\epsilon$) $\kappa<<-\epsilon/8$ tend to split the channel by merging with the main separatrix (Fig.~\ref{f5}a). For $-\epsilon/8\lesssim\kappa\lesssim\epsilon/8$ no off-axis critical points exists (Fig.~\ref{f5}c) and for $\kappa\gtrsim\epsilon/8$, the on-axis hyperbolic point changes to an elliptic one creating a second positive channel (Fig.~\ref{f5}d).

\begin{figure}[h]
 \centering
 \includegraphics[width=0.45\textwidth]{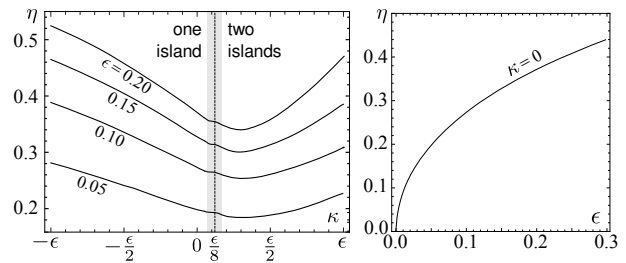}
 \caption{\label{f6} Change in the anisotropy $\eta$ as a function of the ellipticity $\kappa$ for different values of $\epsilon$ (left). Intrinsic anisotropy due to the toroidicity with zero ellipticity (right).}
\end{figure}

Since the ellipticity $\kappa$, and inverse aspect ratio $\epsilon$, define the island sizes and the equilibrium topology, they must be related to the anisotropy $\eta$, in a continuous way. This is depicted in Fig.~\ref{f6}. Increasing the vertical elongation of the current density leads to a growth of the anisotropy of the current channels. The same is not true for the horizontal elongation, since the minimum $\eta$ does not correspond to a vanishing ellipticity. This is due to the toroidicity (represented by $\epsilon$), that adds an implicit anisotropy to the system. For this local solution the anisotropy only vanishes for the cylindrical case ($\epsilon=0,\kappa=0$) and never becomes negative, so it verifies $I_+>2I_-$. The value of $\eta$ saturates during the bifurcation at $\kappa=\epsilon/8$ where, briefly, a zero current island is created. If we consider a {\it very} small region of interest ($\epsilon\rightarrow 0$), the current ellipticity dominates and the equilibrium always presents two islands. 
This means that, if a quasi-stationary transition to an internal reversed current density is possible, the magnetic axis is simultaneously split into three separated axes.

In summary, the magnetic topology related to current density reversals defines several current channels within the plasma. The ratio between the current in the positive channels and the central negative current depends on a topological parameter measuring the anisotropy of the positive channels. In general terms the positive current is about twice the size of the central negative current. This anisotropy was shown to be related to the geometrical properties of the equilibrium in a region of interest inside the plasma. Then, it was observed that the anisotropy is always positive, indicating that the positive channels have {\it at least} twice the current magnitude of the negative one, causing the screening of this channel and forming a structure with net positive current. Finally, it was also shown that experimental magnetic pitch measures in a {\it current hole} are compatible with the existence of a small negative value of the current density.

This work was partially supported by Conselho Nacional de Desenvolvimento Científico e Tecnológico and Fundação de Amparo à Pesquisa do Estado de São Paulo.

\bibliography{references}

\end{document}